\documentclass[12pt]{article}
\usepackage[margin=1in]{geometry}
\usepackage{mathptmx}
\usepackage{mathtools}
\usepackage{amsthm}
\usepackage{natbib}
\usepackage{amssymb}

\usepackage{graphicx}
\usepackage{hyperref, epigraph}
\usepackage{tikz}
	\usetikzlibrary{decorations.pathreplacing}
	\usetikzlibrary{decorations.pathmorphing}
	\usetikzlibrary{patterns}
\usepackage{caption}

\usepackage{etoolbox}
    \makeatletter
    \newlength\epitextskip
    \pretocmd{\@epitext}{\em}{}{}
    \apptocmd{\@epitext}{\em}{}{}
    \patchcmd{\epigraph}{\@epitext{#1}\\}{\@epitext{#1}\\[\epitextskip]}{}{}
    \makeatother

\newcommand{\cd}{\cdot}
\newcommand{\ra}{\rightarrow}
\newcommand{\pr}{\prime}
\newcommand{\de}{\partial}

\newcommand{\abs}[1]{\left\lvert #1 \right\rvert}

\theoremstyle{definition}

\newtheorem*{defin*}{Definition}
\theoremstyle{remark}

\begin{document}

\title{Prospecting a Possible Quadratic Wormhole \\ Between Quantum Mechanics and Plurality}

\author{  Michal Fabinger\thanks{Acalonia Research: michal@acalonia.com.} \and Michael H. Freedman\thanks{Microsoft Research Station Q: michaelf@microsoft.com.} \and E. Glen Weyl\thanks{Microsoft Research Special Projects and RadicalxChange Foundation: glenweyl@microsoft.com.}
}


%

\maketitle

\begin{abstract}
We illustrate some formal symmetries between Quadratic Funding \citep{bhw}, a mechanism for the (approximately optimal) determination of public good funding levels, and the \citet{born} rule in Quantum Mechanics, which converts the wave representation into a probability distribution, through a bridging formulation we call “Quantum Quartic Finance”.  We suggest further directions for investigating the practical utility of these symmetries.  We discuss potential interpretations in greater depth in a companion blog post.
\end{abstract}

\section{Introduction}

In parallel with this paper, we are releasing a blog post\footnote{Available at https://www.radicalxchange.org/media/blog/quantum-mechanics-and-plurality/.}, reflecting on the results here, which we therefore motivate only to the degree necessary to understand the formal statements. This piece results from two convergent motivations.  

One of us recently blogged \citep{freedman22} about a mechanism for public goods funding, Quadratic Funding (QF), proposed by another of us \citep{bhw}, based on the square of the sum of square roots of individual funding contributions.  A friend, Adam Brown, asked if there was a connection to the \citet{born} rule in Quantum Mechanics (QM), where a similar procedure is applied to the complex \emph{amplitudes} (viz. probability amplitudes) to derive the observable probabilities of events.   Concurrently, \cite{weylpluralism, DeSoc} has been exploring, under the banner of ``Plurality'', the formal duality between individuals and groups (e.g. how each can be modeled as arising from the interaction of several units of the other).
 
 In this piece we make a somewhat playful foray into bridging these two distant worlds. In particular, we wonder whether QM might be formally derived from and thus interpreted as a result of a collective compromise among ``agents'' in the spirit of QF, or less ambitiously, if there is a kind of calculation in QM that would map to similar calculations in QF.  While our inquiry here raises far more questions than it settles, we formulate an argument for why such a ``wormhole'' (connection) could exist.

\section{Definitions}

\subsection{What is QF?\nopunct}
QF is an answer to a question: How should public goods be financed in an economy where most goods are private and can be bought with money? A ``public good,'' such as national defense or clean air, is a good (or service) from which members of society naturally benefit and where excluding some individuals from the benefits would be wasteful. Improving the good or increasing its amount is then beneficial for everyone, not just those who made contributions towards the good. Below is a summary of the model presented in \cite{bhw}.

Society consists of $n$ citizens (members) distinguished by an index $i$ with $i = 1,...,n$.
The set of possible public goods is $P$ and we distinguish them by labels $p \in P$. Let $V_i^p(F^p)$ be the currency-equivalent utility citizen $i$ receives if the public funding level of good $p$ is $F^p$. 
These functions are assumed to be smooth and monotone increasing. Total social value is derived as the sum of individual utilities. Issues of partial information and timing of decision making are ignored within the model. The contribution of citizen $i$ towards good $p$ is $c_i^p$. Each contribution is non-negative (a non-trivial assumption). The contribution levels and funding levels form vectors $c$ and $F$, respectively. Let $\mathcal{C}$ and $\mathcal{F}$ denote the vector spaces holding the vectors $c$ and $F$.

\subsubsection{The problem considered}
Find the funding mechanism $\Phi: \mathcal{C} \ra \mathcal{F}$ that maps contribution vectors to funding vectors and maximizes the total social welfare
\begin{equation}\label{eq:welfare}
    W = \sum_{p} \left((\ \sum_{i} V_i^p(F^p)\ ) - F^p\right)
\end{equation}
or all such mechanisms if there are several.

In the absence of external resources, taxation is required to fund the difference between the total of the funding levels and the total of the contributions and it affects citizens' utility levels. Taxation is discussed in section 4.5 of \cite{bhw}, but is not essential here and will be neglected.

\subsubsection{Optimality condition}
Differentiating the social welfare $W$ with respect to the individual contributions $c_i^p$, one finds that for each good $p$,
\begin{equation}
    \sum_i {V_i^p}^\pr (F^p) = 1
\end{equation}
provided the sum is positive at zero funding levels. In defining the model, we specify that the funding level $F^p$ of good $p$ is computed from the individual contributions as
\begin{equation}
    F^p = g\left(\sum_j h(c_j^p)\right)
\end{equation}
We refer to the functions $h$ and $g$, as the weighting function and funding lever, respectively.
The utility of individual $i$ is
\begin{equation}
    U_i^p = V_i^p(F^p) - c_i^p = V_i^p\left(g\left(\sum_j h(c_j^p)\right)\right) - c_i^p
\end{equation}
where the model presumes that the functions $F^p$ are built as indicated from the internal functions $h$ and $g$, the weighting function and funding lever, respectively. Upon differentiating and setting $\frac{\de U_i^p}{\de c_i^p} = 0$, we obtain:
\begin{equation}
    \frac{d V_i^p}{d g} \cd \frac{d g}{d h(c_i^p)} \cd \frac{d h(c_i^p)}{d c_i^p} = 1
\end{equation}

So, the optimality conditions imply for each good $p$:
\begin{equation}
    1 = \sum_i \left(V_i^p\right)^\pr = \sum_i \left(\sum_p \frac{d g}{d h(c_i^p)} \cd \frac{d h(c_i^p)}{d c_i^p}\right)^{-1}
\end{equation}

The possible mechanisms $\Phi$ are democratic in that the funding function $F$ is symmetric in its $n$ variables. The question is to find $g$ and $h$ maximizing social utility. We denote the solution $\Phi^{\mathrm{QF}}$.

\subsubsection{Definition of QF}
The QF mechanism corresponds to
\begin{equation}
\Phi^{\mathrm{QF}}(\{c_i^p\}) = \left(\sum_i (c_i^p)^{\frac{1}{2}}\right)^2
\end{equation}
That is, for every good $p$ its level of funding is the square of the sum of the square roots of the individual contributions.

It was shown in \cite{bhw} that this form of $\Phi^{\mathrm{QF}}$ corresponds to a maximum of social utility and in \cite{freedman22} that this maximum is unique.

\subsection{What is QM?\nopunct}
We will use the term Quantum Mechanics (QM) in a general sense that includes Quantum Field Theory (QFT) and related theories built on the same principles.

\subsubsection{Unitarity axiom}
In QM a (pure) state of a system, possibly even of the universe as a whole, is described by a state vector $\Psi$, an element of a Hilbert space $\in \mathcal{H}$. The state evolves in time via the Schr\"odinger equation $\frac{d\Psi}{dt} = -iH\Psi$, where $-iH$ is an infinitesimal unitary rotation of the Hilbert space, specified by a Hermitian operator $H$. 


\subsubsection{Measurement axiom}
If a measurement is performed, that measurement is characterized by another Hermitian operator $A$. Making the simplifying assumption that $A$ has discrete spectrum, its eigenvectors decompose $\mathcal{H}$ into orthogonal eigenspaces. Using such a decomposition, the state vector may be written as $\Psi = \sum_i a_i \Psi_i$. The measurement of $A$ projects $\Psi \in \mathcal{H}$ into any one of these components. The probability $p_i$ of observing the $i$th eigenvalue of $A$ is given by the Born rule: $p_i = \abs{a_i}^2 \abs{\Psi_i}^2\abs{\Psi}^{-2}$. That is, the measurement outcome is always an eigenstate of the observable $A$, and the probability of the $i$th outcome is determined by the norm squared of the mysterious ``probability amplitude,'' the complex coefficient $a_i$. (If one follows the convention that, at all times, state vectors $\Psi$ and $\Psi_i$ are normalized to have norm one, then  $p_i = \abs{a_i}^2$ .)

The modern reader may be surprised that in \citet{vNM}'s treatise on QM the measurement axiom corresponds to his axiom 1 and the unitarity axiom to his axiom 2. Today only \emph{unitarity} is regarded as fundamental and the \emph{measurement axiom} is regarded as a high-fidelity approximation to a ``decoherence'' process interacting system and environment, experiment and apparatus, which at root only involves the unitarity axiom. The measurement axiom is, in effect, relegated to a shorthand notation, for a unitary evolution which results in a ``final'' state which, up to tiny errors, can be written as a sum of orthogonal terms: $\sum_i a_i \vert \Psi_i\rangle \otimes \vert \mathcal{E}_i\rangle$. Here $\vert \Psi_i\rangle$ is a ``system'' state, what is being measured, and $\vert \mathcal{E}_i\rangle$ is an environmental ``pointer state'' which reads off the measurement outcome. Again, the Born rule tells us that the chances of finding the system in the $i$th state is $\abs{a_i}^2$.

The perspective closest to the one we sketch here has been developed by \citet{zurek09,zqz09} under the name \emph{Quantum Darwinism} (QD). \citeauthor{zurek09} observes that an experimenter (human or otherwise) only samples a tiny random piece of the environment. In his view a state becoming \emph{collapsed}, that is becoming macroscopically correlated, involves an ``evolutionary struggle'' to copy its information widely into numerous small fragments of the environment.  Since the universe cannot record the entire past and still do new things, being recorded/replicated is a limited resource and drives this microscale Darwinism.\footnote{Note that copying eigenstates of physical observables, unlike copying the general pure state, does not violate the No-Cloning Theorem.}

While QD sees the Born rule as emerging from an individualistic struggle of states to leave a record of themselves, the study here searches for a mechanism \emph{quantum quartic finance} (Q4F) where virtual processes cooperate with the common goal of leaving some mark behind, that is entangling with macroscopically many degrees of freedom, which, generically, constitutes irreversible decoherence. Unlike the weighted contributions of QF, amplitudes have a phase that leads to the \emph{4th power} being taken in Q4F; the function $g$ in Q4F will turn out to be $g(x)= a x^4$, where $a$ is a positive constant determined below. 

Our goal is to understand the analogy between \emph{norm-squared of the sum of amplitudes} in quantum mechanics and the \emph{square of the sum of weighted contributions} in quadratic finance. In pursuit of this analogy, we elevate virtual processes, VP, to be the ``citizens'' of Q4F, and their amplitudes to be their ``weighted contributions'' since we know from QM that their amplitudes must be added up, and if measured, the norm squared of the sum determines the observation probability. The Q4F model below will identify the funding function $F^p$ with the {\it square} of that probability. The appearance of a fourth power may be seen as compensating for the ubiquitous square-root behavior associated to random walks. 

\subsection{Interpretation}
We will shortly build and analyze a stochastic mathematical model whose solution is Q4F. Our fanciful intention is to interpret this model in the context where the $j$th citizen is a virtual process, VP, $j$. For example, $j$ might be a photon emitted by a source and passing through one particular slit, in the double slit experiment, on its way to a detector. Or it might be the possibility of a change in electron state of atom A that could release a photon which could then be absorbed by atom B changing B's state. It could be a single ``propagator'', a virtual exchange within a larger Feynman diagram (FD), or an entire FD. 
\footnote{In QFT propagators are ``dressed'' by interaction with the vacuum, so the distinction between individual pieces of a diagram and entire diagrams is not tenable, and we do not make it. Normally one thinks of a continuum of FD realizations as their vertices are parameterized by location in physical or momentum space, so that each diagrammatic form is evaluated therefore by a finite dimensional integral over copies of space. One may presume a finite regularization, and treat the number of our citizens as finite, and decide not to worry about the familiar difficulties of taking limits encountered in constructive field theory. Or alternatively allow infinitely many citizens and replace later summations below by integrals. The discussion here is only at a formal level so either choice is fine.}

We assume each citizen $j$ possesses some resource (currency) from which it can contribute $c_j^p$ to good $p$. The goods $p$ (in Q4F) are quantum events that, as in Zurek's conception, have the potential to cascade into a kind of immortality---a dark spot on a film or a record in a detector, or merely entanglement with any macroscopic bit of matter regardless of whether a human is intentionally performing an experiment. The metaphor is that each virtual process wants to leave a record of itself in the universe, to be measured, to become entangled with a macroscopic state, to decohere.

\subsubsection{Definition of Q4F}
The Q4F mechanism corresponds to
\begin{equation}
    \Phi^{\mathrm{Q4F}}(\{c_i^p\}) = a \left(\sum_{i=1}^n \omega_i^p (c_i^p)^{\frac{1}{4}}\right)^4
\end{equation}with the constant $a$ determined below. Here $n$ is the number of citizens $i$ and the phase factor $\omega$ is a unit complex number that follows a uniform distribution. For every good $p$, a potentially entangling event, the probability of achieving good $p$ is the square of the sum of phased fourth roots of the individual contributions. 

In actual QM each VP has a phase which can be worked out (up to an overall indeterminacy), but as in Wigner's random matrix theory, when a large number of $i$ are contributing to an aggregate amplitude it may be a reasonable approximation to regard the individual contributions as random, and indeed, if the model is attempting to reflect ``intentionality'' on the part of the VP, the VP will have no information about the phase of other contributing VPs so it cannot do better---in terms of increasing the probability of $p$ than by contributing a randomly phased fourth root to the sum.

The funding mechanism of Q4F has become a rule, the Born rule, for aggregating virtual processes into a kind of probability \emph{squared} of an event (a ``project'') $p$. There are two essential differences between QF and Q4F which we will see below precisely cancel each other, on passing to expectation values. The first difference is that to model quantum mechanical phase, when passing from $i$'s currency (not defined here) to its amplitude, we supplied a phase factor $\omega$ to its fourth root. Then, because phase factors statically produce extensive cancellations---recall that a random walker after n steps in only at distance about $n^{\frac{1}{2}}$ from her starting point---the funding/probability formula employs a quartic $g$. Under our assumption of random phase the Q4F formula, similarly to \cite{freedman22}, is likely optimal (in expectation) with similar model assumption as in QF (chiefly linearity and monotonicity of the $V_i^p$); but we only carried the calculation out assuming that $g(x)=ax^r$ and $h(y)=by^{\frac{1}{q}}$, for some $r,q>0$.

\section{Potential Next Probes}

If the ‘wormhole’ we explore is usable, it should be possible to pass some technique from quantum physics into economics or social science or possibly the reverse.\footnote{Of course, many proposed connections between physical and social sciences are superficial, and this one could be, too. But some
connections of this kind turned out to be useful, which motivates the speculation here.} In condensed matter physics Hamiltonians are sorted into $k$-body terms $k= 1,2,3,\dots$
 Across the wormhole can we see such terms as communities? Since Wigner’s study of heavy nuclei, it has been customary to approximate complicated Hamiltonians as random within their symmetry ensemble. This approximation explains the wide application of his “semicircle law” (the spectral distribution of random Hamiltonians); is there an equivalent semicircle law of social science? 

Just as individuals together constitute communities and polities, one of us \citep{crossroads} has emphasized the inverse relation whereby the individual emerges from the intersection of the groups in which they participate. This is reminiscent of the dual perspective: operators and states. The states may be thought of as functional on operators or the operators as functionals on state. In quantum mechanics this leads to the Heisenberg and Schrodinger pictures, respectively.  A natural line of inquiry, then, is whether these dual perspectives can be translated to the social sciences. Because the duality is a structural formality in physics, it is rare that one perspective or the other is more powerful. But the duality became a powerful tool in the hands of \citet{dyson} in which the main term is treated in the Heisenberg picture and only the perturbation in the Schrodinger picture. This ‘Dyson picture’ marvelously isolates the effect of the perturbation. Can such a technique be pulled through the wormhole to facilitate perturbation theory in economics and sociology?\footnote{We would not expect the techniques to be literally the same in social science, just analogous.} 

Because the groups that individuals participate in differ dramatically in scale (from friend groups to cities and nations) are often usefully seen hierarchically, with the world being constituted at least partly by integrated groupings of communities into large polities.  This is echoed on the physical side in the renormalization group which studies the behavior of physical systems when viewed at larger scales and lower energies. A great success of this program is the concept of a low-energy effective field theory. It has taught us to expect that even the basic, descriptive, language must be altered as one descends in energy: In a quantum Hall fluid, quasi-particles, not electrons, become the fundamental particles. Can renormalization be applied to the relationships among communities to form larger societies?\footnote{This would correspond to a model with fewer agents capturing the main features of a more detailed model that has more agents.}

A final puzzle in QF that the connection to QM seems relevant to resolve is the interpretation of negative and imaginary numbers.  In QF, negative contributions are sometimes allowed, but sign conventions are twisted to make these work and there has been, to our knowledge, no attempt to seriously grapple with the proper interpretation of a negative funding level, perhaps a decentralized means of punishment of harmful activities.  Complex phase factors are central to QM.  Perhaps there is a way to draw on the formalism of QM to allow a more principled and broadly applicable treatment of negative and imaginary contributions, funding levels and intermediate ``votes''.

\section{The model}

We describe a model whose solution is Q4F. Here we consider the problem in greater generality: instead of random complex phases, we consider random direction vectors uniformly distributed over a $(d-1)$-dimensional sphere, with $d\geq 1$ For $d=2$, this is equivalent to the case of the complex phases. For $d=1$, this is equivalent to the case of two possible phase factors: $+1$ and $-1$.

We add this generality for two reasons: (1) QM may be defined over $\mathbb{R}$, $\mathbb{C}$, or $\mathbb{H}$ (the quaternions) leading to unit 0-, 1-, and 3-spheres, and (2) ignoring QM, the target spaces $S^{d-1}$ correspond to well studies spin models in which statistical variance decreases with th dimension $d$ (the so-called orthogonality catastrophe).

We denote the contribution from individual $i$ towards project $p$ as
$$c_i^p$$
The random direction associated with contribution from individual $i$ towards project $p$ will be denoted
 $$\omega_i^p \in S^{d-1}$$
 Here $S^{d-1}$ is a unit sphere in a $d$-dimensional space. In the case of $d=2$ each two-component vector $\omega_i^p$ may be identified with a complex number of absolute value 1. In the case of $d=1$, $S^0$ corresponds to the set of two numbers $\{-1,1\}$.
 
  We denote the list of the random directions for all the individuals as
 $$\omega^p = (\omega_1^p,\omega_2^p,\dots,\omega_n^p)$$
The funding amount for project $p$ given the values of the random directions is assumed to be of the form
$$\tilde F_{\omega}^p = g(\lVert \sum_{i=1}^n \omega_i^p h(c_i^p) \rVert)$$
 Here $\lVert \cdot \rVert$ stands for the L2 norm of a $d$-dimensional vector; simply the Euclidean length of that vector.
 
 The utility of individual $i$ associated with project $p$ is
 $$\tilde U_i^p = V_i^p(\tilde F_{\omega}^p)-c_i^p$$
  Similarly, the expected utility of agent $i$ associated with project $p$ is
  $$E(\tilde U_i^p) = E(V_i^p(\tilde F_{\omega}^p))-c_i^p$$
  The derivative of the expected utility with respect to $c_i^p$ is
  $$\frac{\partial}{\partial c_i^p}E(\tilde U_i^p) = \frac{\partial}{\partial c_i^p}E(V_i^p(\tilde F_{\omega}^p))-1$$
  If the function $V_i^p$ is close to linear over the range of typical funding values, we have, 
  approximately\footnote{Note that the function still needs to be sufficiently non-linear so 
  that we can consider achieving social optimality by satisfying the condition  $ \sum_{i=1}^n {V_i^p}' =1 $  below.}
 $$\frac{\partial}{\partial c_i^p}E(\tilde U_i^p)\ =\ {V_i^p}' \frac{\partial}{\partial c_i^p} E(( \tilde F_{\omega}^p))-1$$
 The approximation is more accurate if the funding level is less volatile (which happens for larger $d$).

We will assume the following functional form 
 $$g(x) = a y^r$$
  with $r$ a positive real number.
 
 Then
  $$\tilde F_{\omega}^p =g(\lVert \sum_{i=1}^n \omega_i^p h(c_i^p) \rVert) = a (\lVert \sum_{i=1}^n \omega_i^p  {h(c_i^p)} \rVert)^r$$
 For the expectation, we have
  $$E(\tilde F_{\omega}^p) =E(g(\lVert \sum_{i=1}^n \omega_i^p h(c_i^p) \rVert)) = a E((\lVert \sum_{i=1}^n \omega_i^p  {h(c_i^p)} \rVert)^r)$$
  To keep the discussion relatively simple, we will choose the value $r=4$ and show how to choose other parameters of the functions so that the individual and social optimality conditions are identically satisfied. It is possible to show that other values of $r$ would not allow for this joint optimality. The resulting funding mechanism is
$$\tilde F_{\omega}^p = a(\lVert \sum_{i=1}^n \omega_i^p h(c_i^p) \rVert)^4$$

We evaluate $E(( \tilde F_{\omega}^p))$ under the assumption that $n-1$ values of the contributions are sufficiently uniform so that we can use the law of large numbers for these $n-1$ contributions. At this point we do not impose any similar assumption the one remaining contribution. We allow for the possibility of this contribution being very large. We will use index $i$ for this potentially contribution here and indices $j$ for the other $n-1$ contributions.

The details of the evaluation are included separately below\footnote{See 
"Derivation of the expected value" below.}, and the result of the evaluation is\footnote{Note that this formula does not exhibit a symmetry between the $i$th contribution and $j$th contributions for $j\neq i$. This is because we applied a central-limit-theorem-based approximation to $j$th contributions with $j\neq i$, which we assumed to be and sufficiently uniform. The $i$th contribution was singled out and could be potentially very large. If the $i$th individual behaves similarly to the other individuals, the $i$th contribution will not be excessively large, and we can ask how the result could compare to applying the central limit theorem directly to all terms. We find that the formula here gives a different result than the central limit theorem applied directly to all $n$ contributions, which could give a symmetric expression. But there is no contradiction because the discrepancy divided by the overall size goes to zero as $n\rightarrow \infty$.}
 $$ E(( \tilde F_{\omega}^p))=a\  \left(\frac{d+2}{d}  (  \sum_{j=1,j\neq i}^{n}  {h(c_j^p)^2})^2 +  \frac{2 (d+2)}{d}  \ h(c_i^p)^2 ( \sum_{j=1,j\neq i}^{n}  {h(c_j^p)^2}) + h(c_i^p)^4\right) $$
  Therefore
  $$\frac{\partial}{\partial c_i^p} E(( \tilde F_{\omega}^p))=a\ h'(c_i^p) \left(\frac{4 (d+2)}{d}  h(c_i^p)  ( \sum_{j=1,j\neq i}^{n}  {h(c_j^p)^2}) +4 (h(c_i^p))^3\right) $$
  If $c_i^p$ is comparable to the typical contributions from other agents, we can approximately write
  $$\frac{\partial}{\partial c_i^p} E(( \tilde F_{\omega}^p))= \frac{4 (d+2)}{d}a\ h'(c_i^p)   h(c_i^p)   \sum_{j=1}^{n}  {h(c_j^p)^2}  $$
  The first-order condition of individual $i$ is 
 $${V_i^p}' E((\frac{\partial}{\partial c_i^p} \tilde F_{\omega}^p))-1 =0$$
 $${V_i^p}' \ \frac{4 (d+2)}{d}a\ h'(c_i^p)   h(c_i^p)   \sum_{j=1}^{n}  {h(c_j^p) ^2}  =1$$
  which gives
 $${V_i^p}' = \frac{d}{4(d+2)} \frac{1}{a h'(c_i^p)   h(c_i^p)   \sum_{j=1}^{n}  {h(c_j^p)^2}} $$
 For social optimality we need
  $$\sum_{i=1}^n {V_i^p}' =1 $$
  Substituting for ${V_i^p}'$ leads to the condition
  $$\sum_{i=1}^n  \frac{d}{4(d+2)} \frac{1}{a h'(c_i^p)   h(c_i^p)   \sum_{j=1}^{n}  {h(c_j^p)^2}} =1$$
 Now we specify a power-law functional form for the function $h$:
  $$h(y) = b y^{\frac{1}{q}}$$
  For this functional form, the social optimality condition becomes
  $$\sum_{i=1}^n  \frac{d}{4(d+2)} \frac{1}{a \frac{b}{q}  (c_i^p)^{\frac{1}{q}-1}   b (c_i^p)^{\frac{1}{q}}    \sum_{j=1}^{n}  {(b (c_j^p)^{\frac{1}{q}})^2}} =1$$
  $$\sum_{i=1}^n  \frac{d}{4(d+2)} \frac{q}{a b^4} \frac{1}{  (c_i^p)^{\frac{2}{q}-1}       \sum_{j=1}^{n}  {(c_j^p)^{\frac{2}{q}}}} =1$$
  $$\frac{d}{4(d+2)} \frac{q}{a b^4} \sum_{i=1}^n (c_i^p)^{1-\frac{2}{q}}      =\sum_{j=1}^{n}  {(c_j^p)^{\frac{2}{q}}}$$
  We see that we need to choose
  $$q=4$$
  $$ab^4 = \frac{d}{d+2} $$
  Then the social optimality becomes simply
 $$\sum_{i=1}^n (c_i^p)^{\frac{1}{2}}      =\sum_{j=1}^{n}  {(c_j^p)^{\frac{1}{2}}}$$
  which is identically satisfied. (If we chose any other $r$ besides $r=4$, we would not have been able to get the resulting equation identically satisfied.)
  We conclude that the funding mechanism corresponds to
\begin{equation}
    \Phi(\{c_i^p\}) = \frac{d}{d+2} \left(\sum_{i=1}^n \omega_i^p (c_i^p)^{\frac{1}{4}}\right)^4
\end{equation}
Specifically for $d=2$, this gives the result
\begin{equation}
    \Phi^{\mathrm{Q4F}}(\{c_i^p\}) = \frac{1}{2} \left(\sum_{i=1}^n \omega_i^p (c_i^p)^{\frac{1}{4}}\right)^4
\end{equation}
  
  For a more detailed understanding, note that the distribution of\footnote{Here we suppressed the index $p$ for simplicity of notation.} 
  $$(\lVert \sum_{i=1}^n \omega_i  {h_i} \rVert)^4$$
is approximately the distribution of the fourth power of the length of a multivariate normal distribution in $d$ dimensions with mean zero and a diagonal variance-covariance matrix with all diagonal entries equal to
 $$\sigma^2=\frac{1}{d}  \sum_{i=1}^{n}  {h_i^2} $$
  For such a multivariate normal distribution, the PDF of the distance from the origin $\tilde r$ is
  $$p\left(\tilde r\right)=\frac{2 \pi^{\frac{d}{2} }}{\Gamma\left(\frac{d}{2}\right)\sigma} \left(\frac{\tilde r}{\sigma} \right)^{d-1}\exp\left(-\frac{1}{2}\left(\frac{\tilde r}{\sigma} \right)^{2}\right)$$
 where $\Gamma$ is the gamma function.
  Having this PDF makes it possible to straightforwardly derive the PDF of the funding level and different characteristics of that distribution.
  As $d\rightarrow \infty$, the ratio of standard deviation to mean of the funding level approaches zero, and there is no uncertainty. In this limit, we get the same contribution levels and funding levels as for quadratic funding. In other words, as $d\rightarrow \infty$, quartic finance becomes exactly equivalent to quadratic finance.

The equivalence of the quartic model in the $d\rightarrow\infty$ limit and QF may not be obvious, so let us elaborate on it. Let us consider a model with a specific value of $d$ and look at the expected value $E( \tilde F_{\omega}^p)$ of the funding for project $p$. The formula at the end of the last section implies\footnote{Here we only need an expression under the assumption that all $c_i^p$ are sufficiently uniform and none of them is especially extreme so that the central limit theorem applies to the contributions from all individuals. Technically, we get the implication by considering the formula at the end of the next section, setting $h(c_n^p)=0$ to disregard the $n$th contribution, and then replacing $n$ in the formula by $n+1$ so that the sum has $n$ terms.}
$$ E( \tilde F_{\omega}^p)=a\  \frac{d+2}{d}  (  \sum_{j=1}^{n}  {h(c_j^p)^2})^2  $$
Substituting $h(c_j^p)=b\ (c_j^p)^{1/4}$ and $a=b^{-4}d/(d+2)$ gives
$$ E( \tilde F_{\omega}^p)= (  \sum_{j=1}^{n}  (c_j^p)^{\frac{1}{2}})^2  $$
The right-hand side coincides with the funding prescription in QF. The difference is that here we have randomness in funding even for fixed contributions from the individuals. This randomness comes from the random directions $\omega_i^p$. With increasing $d$, however, the probability distribution of the total funding level becomes more and more concentrated around the expected value, as stated above. In the $d\rightarrow\infty$ limit, the probability distribution places 100\% probability on the this value. In other words, in the $d\rightarrow\infty$ limit, the model becomes exactly the same as QF.

 \subsection{Derivation of the expected value}
 Here we will calculate 
  $$E((\lVert \sum_{i=1}^n \omega_i  {h_i} \rVert)^4)$$
  for a vector $h$, where one of its entries is allowed to be potentially very large. We used the result of this calculation in the discussion above. In the case of project $p$, we will use $h_i = h(c_i^p)$. We suppress the project index $p$ for the directions $\omega_i$ in this calculation.
 
 Let us assume that $h_1, h_2, ... h_{n-1}$ have values for which we can use the central limit theorem, that is, values that are sufficiently uniform. But at this moment we will not assume that $h_n$ has a comparable size. We allow for the possibility that the term with $h_n$ dominates the sum.
 
 We have

$$E((\lVert \sum_{i=1}^n \omega_i  {h_i} \rVert)^4)=$$

$$E((\lVert \sum_{i=1}^{n-1}  \omega_i  {h_i} \rVert^2 +2 (\sum_{i=1}^{n-1}  \omega_i  {h_i}) \cdot \omega_n  h_n + h_n^2 )^2)$$
 
 Since the distribution of the directions is uniform, 

$$E((\lVert \sum_{i=1}^n \omega_i  {h_i} \rVert)^4)=$$

$$E(\lVert \sum_{i=1}^{n-1} \omega_i  {h_i} \rVert^4  )+ 2 h_n^2 E( \lVert \sum_{i=1}^{n-1} \omega_i  {h_i} \rVert^2   )+4 h_n^2 E( (\sum_{i=1}^{n-1} \omega_i  {h_i})\cdot \omega_n   )^2)  + h_n^4 $$
 By the central limit theorem (or its generalization), $\sum_{i=1}^{n-1} \omega_i  {h_i} $ approximately follows a multivariate normal distribution with mean zero and a diagonal variance-covariance matrix with all diagonal entries equal to
$$\sigma^2=\frac{1}{d}  \sum_{i=1}^{n-1}  {h_i^2} $$
For the first term, we need the following intermediate results
 $$E(\lVert \sum_{i=1}^{n-1} \omega_i  {h_i} \rVert^4  )\approx d\ \kappa\ \sigma^4  + d(d-1)\sigma^4 $$
  Here $\kappa$ is the kurtosis of the normal distribution, namely $\kappa=3$. We can write this intermediate result more simply as
  $$E(\lVert \sum_{i=1}^{n-1} \omega_i  {h_i} \rVert^4  )\approx d(d+2)\sigma^4$$
  For the second term, we need 
 $$E( \lVert \sum_{i=1}^{n-1} \omega_i  {h_i} \rVert^2   )\approx d\ \sigma^2 $$
  For the third term, we need 
 $$E( (\sum_{i=1}^{n-1} \omega_i  {h_i})\cdot \omega_n   )^2) =\sigma^2$$
 Combining the terms, we get
 $$E((\lVert \sum_{i=1}^n \omega_i  {h_i} \rVert)^4)\approx d(d+2)\sigma^4 + 2 (d+2)\ h_n^2 \sigma^2 + h_n^4$$
  Recalling the definition $\sigma^2=\frac{1}{d}  \sum_{i=1}^{n-1}  {h_i^2} $, we conclude that 
  $$E((\lVert \sum_{i=1}^n \omega_i  {h_i} \rVert)^4) \approx \frac{d+2}{d}  (  \sum_{i=1}^{n-1}  {h_i^2})^2 +  \frac{2 (d+2)}{d}  \ h_n^2 ( \sum_{i=1}^{n-1}  {h_i^2}) + h_n^4$$
  For the expected value of the funding, this implies
 $$ E( \tilde F_{\omega}^p)=a\  \left(\frac{d+2}{d}  (  \sum_{i=1}^{n-1}  {h(c_i^p)^2})^2 +  \frac{2 (d+2)}{d}  \ h(c_n^p)^2 ( \sum_{i=1}^{n-1}  {h(c_i^p)^2}) + h(c_n^p)^4\right) $$
  In this derivation we singled out, without loss of generality, the $n$-th contribution. If we formally change the labels (indexes) of the contributions so that instead it is the $i$-th contribution that is singled out (allowing it to be arbitrarily large), we get the formula
  $$ E( \tilde F_{\omega}^p)=a\  \left(\frac{d+2}{d}  (  \sum_{j=1,j\neq i}^{n}  {h(c_j^p)^2})^2 +  \frac{2 (d+2)}{d}  \ h(c_i^p)^2 ( \sum_{j=1,j\neq i}^{n}  {h(c_j^p)^2}) + h(c_i^p)^4\right) $$

\bibliography{arxiv}

\end{document}